\documentstyle[12pt,aasms4]{article}

\begin{document}

\title{Interactions Between Solar Neutrinos and
Solar Magnetic Fields}
\author{D. S. Oakley}
\affil{Colorado Christian University, Lakewood, CO 80226}
\authoremail{oak@colorado.edu}

\and

\author{Herschel B. Snodgrass}
\affil{Physics Department, Lewis and Clark College, Portland, OR 97219}
\authoremail{hbs@lclark.edu}

\begin{abstract}
 
We attempt to correlate all of the available
solar-neutrino capture-rate data with the strong
magnetic fields these neutrinos encounter
in the solar interior along their Earth-bound path.
We approximate these fields using the
(photospheric, magnetograph-measured)
surface magnetic flux from central latitude bands,
time delayed to proxy the solar interior.
Our strongest evidence for anticorrelation
is for magnetic fields within the central $\pm 5^{0}$ solar-latitude
band that have been delayed by 0.34 years.
Assuming a neutrino-magnetic interaction, this
might indicate that interior fields travel to the solar surface in
this period of time.
As more solar-neutrino flux information is gathered,
the question of whether such evidence is physical or statistical
in nature should be resolved, providing that
new data span enough solar cycles and that correlation studies
focus on these isolated fields.

\end{abstract}

\section{Introduction}
 
The mystery of the missing solar neutrinos, which impacts
our understanding of the Standard (electro-weak) Model, solar magnetic fields,
solar-nuclear processes, and cosmology, continues
to deepen as more data are collected.
Discrepancies between theory and solar-neutrino experiments,
as well as between the experiments themselves,
has lead to investigations of experimental errors, flawed solar
models, and energy-dependent
processes which remove neutrinos before they reach Earth.
For example, despite the $\sim$
tenfold difference between their thresholds of 7.3 and 0.8 MeV, respectively,
Kamiokande and Homestake
find similar capture rates\markcite{dav1} (Davis 1994).
While both of these are still well below that predicted,
experiments sensitive to lower energy thresholds, 
SAGE\markcite{sage} (Abdurashitov et al. 1994) 
and GALLEX\markcite{gall} (Anselmann et al. 1993),
detect rates closer to theoretical predictions.
Barring experimental error this would
indicate a maximum in the `missing neutrino spectrum' near the 1-7 MeV range.
 
Throughout the $25$ years since the neutrino deficit was first noted,
efforts to find experimental error or to adjust solar
models have steadfastly proved unsuccessful.
Thus, increasing attention has been focussed on two possibilities;
a flavor-changing interaction occurs
in the dense solar interior (MSW effect\markcite{msw},
Wolfenstein 1978; Mikeyev \& Smirnov 1985)
and/or a neutrino-magnetic
interaction stronger than that predicted
by the standard model occurs\markcite{mar,lim} 
(Marciano \& Sirlin 1988; Lim \& Marciano 1988).
Both of these might alter the neutrino flux from that
expected in the measurements.
 
The MSW effect shows promise because
calculations, with mass-eigenstate and mixing parameters in the range of
$\sim 10 ^{-5}eV^{2} < \Delta m^{2}< \sim 10^{-7} eV^{2}$
and $0.07<sin^{2} 2\theta < 0.76$, reveal preferential flavor-changing
interactions for solar neutrinos in the 1-7 MeV range\markcite{lang,hata}
(Langacker 1994; Hata \& Langacker 1994)
(flavor oscillations in the
vacuum alone, with the Earth at
a node, could also resolve the experimental discrepancies\markcite{lang} 
(Langacker 1994) if
$\Delta m^{2}\sim 10^{-10} eV^{2}$ and $sin^{2} 2\theta > 0.7$).
In a recent experimental search for neutrino
oscillations\markcite{osc}
(C. Athanassopoulos et al. 1995), however, the
mass-eigenstate difference was inferred
to be on the order of 0.1 eV for these $sin^2 2\theta$ ranges,
which is many orders of magnitude larger
(assuming the same flavors are involved).
The existence of such neutrino-flavor
oscillations are, of course, very uncertain experimentally.
 
The suggestion of a magnetic interaction has grown out of
a number of studies\markcite{dav2,wil}
(see e.g. Davis 1987; Wilson 1987)
that find an anticorrelation between
the Homestake capture rate and indicators of the $\sim$ 11 year solar
magnetic activity cycle.
The effect suggested is that during
periods of high solar-magnetic activity, neutrinos
are effected such that the detected solar-neutrino flux on Earth is low
(and visa-versa).
While the statistical case for such an
anticorrelation is inconclusive, intriguing results have recently emerged
which suggest that
a neutrino-magnetic interaction beyond that predicted by the standard model
(but still within presently observed limits\markcite{oak1,mcnut},
Oakley et al. 1994; McNutt 1995) is possible.
 
In this paper we will attempt to explore such an anticorrelation between the
solar neutrinos and the strong equatorial-magnetic fields they encounter
in the solar interior.
To isolate the magnetic fields near the solar equator,
along the neutrinos' Earth-bound path,
the solar-latitude dependence will be investigated, and in order
to isolate the fields in the solar interior,
the time-delay dependence of the fields will be investigated.
 
While we hope to show evidence that is compelling enough to
warrant further correlation studies, particularly with future
data that include neutrino-flavor information, we also hope to
establish procedures which allow for
correlations with actual magnetic fields the neutrinos encounter.
These procedures may also ultimately
be valuable in investigating solar-magnetic phenomena themselves.
 
\section{Correlation Studies}
 
A solar neutrino anticorrelation with the solar cycle was first
observed by Davis\markcite{dav2}
(Davis 1987), who cross-correlated a
highly-smoothed Homestake
neutrino capture rate against total sunspot number.
A number of subsequent studies
presented weak statistics either in
support\markcite{wil,cho,bieb,dav3,bah}
(Wilson 1987; Choudhuri 1989; Bieber et al. 1990;
Davis \& Cox 1991; Bahcall \& Press 1991)
or refutation\markcite{kam1,morr,shi}
(Hirata et al. 1990; Morrison 1992; Shi et al. 1993), but as the
data continues to roll in, the anticorrelation appears to be becoming
more robust.
It has now been inferred in studies using other solar cycle indicators,
including solar wind\markcite{mcnut}
(McNutt 1995), the p-mode frequency shifts
from helioseismology\markcite{kraus,del}
(Krauss 1990; Delache et al. 1993), and the coronal 
green line\markcite{coro} (Massetti \& Storini 1996).
Since all of these solar-cycle indicators are,
of course, correlated with each other, correlation of the neutrino flux with
any one of them implies correlation with all; thus
the plethora of these results does not
increase our confidence that the correlation is not accidental.
Some studies, however, do offer additional evidence: in correlating the
Homestake capture rate
against the surface magnetic fields, 
for instance, it has been found\markcite{oak1} (Oakley et al. 1994)
that the anticorrelation is significantly stronger
when the fields that are not along the neutrino flight path are
excluded from the calculation, and, although counting statistics from the
other capture experiments are still too small for
reliable comparison, the same result is suggested using
the Kamiokande capture rate.
It should be noted that the SAGE and GALLEX
experiments are
still too new to permit any significant temporal correlation studies.
 
Such magnetic interactions may account for all of the missing
neutrinos and it has been claimed\markcite{mcnut} 
(McNutt 1995) that the
the energy dependence can also be reproduced
using resonance calculations similar to MSW (using magnetic states).
While the neutrino magnetic moment required is
marginally within what is allowed by measurement\markcite{krak,oak1}
(Krakhauer et al. 1990; Oakley et al. 1994), others
have suggested that this hypothesized magnetic interaction
could not, by itself, fully account for the missing neutrinos\markcite{shi}
(Shi et al. 1993).
Confirmation of any neutrino-magnetic
interaction, however, is interesting aside from the `solar neutrino problem'
and would imply that the neutrinos must have a rest mass, which
is not only compatible with MSW but also has a great impact on
cosmological questions concerning big-bang nucleosynthesis
and galaxy formation\markcite{foot}
(Foot \& Volkas 1995).
 
\section{Solar Magnetic Fields}
 
The magnetic field observed
at the solar surface is too weak to have any such effect on the neutrinos and
so the anticorrelation, if it is real,
must come from the fields that lie in the solar interior
at the base of the convection zone.
Certain magnetic buoyancy arguments predict that these fields rise to the
solar surface on the order of a few months\markcite{magb}
(Parker 1975),
possibly longer or shorter if in the presence of large-scale convective motions.
Given the observed limits on neutrino-magnetic moments,
the magnitude of the field required
to describe the neutrino deficit is still uncertain.
Some suggest fields
near $10^5$ Gauss are needed,
though fields as low as 10$^{3}$ Gauss may 
suffice\markcite{mar,lim,mcnut} 
(Marciano \& Sirlin 1988; Lim \& Marciano 1988; McNutt 1995).
Although the existence of fields greater than $10^5$ Gauss
has been inferred from Helioseismology data\markcite{pet,dsil,gou,dzi}
(Petrovay 1991; D'Silva 1992; Gough \& Thompson 1988; Dziembowski \& Goode 1989),
the measurements are both controversial and too uncertain at
present to properly quantify.
 
If we assume that the these deep field structures exist,
are strong enough to effect the neutrinos,
and rise to the solar surface,
then any neutrino-magnetic correlation studies, using photospheric data,
should explore the question of timing.
For example, in the event of neutrino interactions
with interior fields which precede the surface measurements
by one year, then strongest
anticorrelations should be found in a data set where the surface fields are
delayed accordingly, i.e. {\it
a 1985 neutrino would interact with the field proxied
by the 1986 surface}.
 
%%%%%%%%%%%%%%%%%%%%%%%%%%%%%%%%% Experiment
%%%%%%%%%%%%%%%%%%%%%%%%%%%%%%%%%%%
\section{Data Reduction and Correlation Statistics}
 
Here we investigate such time delays using
all available neutrino data (from GALLEX,
SAGE, Kamiokande, and Homestake)
with magnetograph measured solar surface
magnetic fields from Mt. Wilson.
In particular, the Homestake data,
which have a longer running time than any other
solar neutrino experiment,
are correlated against magnetic data using the
Spearman rank-order correlation, which has been used in previous studies
\markcite{bah,oak1} (Bahcall \& Press 1991; Oakley et al. 1994).
The important statistic here,
along with the correlation coefficient itself, is the significance
level attached to it
which gives the probability that correlation against a random
array would give an equal or greater coefficient.
Note, therefore, that  a {\it low} significance level is
indicative of a meaningful correlation,
and by convention a correlation is considered
`highly significant' when the significance
level is less than 1\%.
 
Because we wish to find a correlation on
the time span compatible with the 11 year cycle, the
1/10 year bins used in these previous studies
yielded correlations that were 
unreasonably strong\markcite{new}
(Newman et al. 1994).
Therefore, we will correlate Homestake with magnetic
data that are binned into the irregular Homestake intervals,
ranging from 0.1 to 0.6 year, with no additional smoothing or massaging.
Of course, even though we would get better statistics with finer binnings,
in proportion to the ratio of bin interval/correlation length (11 years),
our larger bin intervals should still yield the same trends, if any exist.
 
It has been suggested that Homestake may also be sensitive to
large solar flares\markcite{dav4} (Davis 1995).
Because flare observations
usually occur during periods of high magnetic activity they would be expected
to decrease the anticorrelation.
There is doubt as to whether the neutrino flux
created by these flares is observable at Homestake and we find that the
elimination of the data intervals that contain the 10 strongest
flares\markcite{flar} 
(McIntosh 1995) causes no significant change in the correlation.
 
The remaining solar neutrino data, from GALLEX, SAGE, and Kamiokande,
do not span enough solar cycles to allow such a rank-order correlation analysis.
Here, however, we can employ scatter plots, regressions, and
explore general trends.
Such techniques have been used
to demonstrate null correlation in previous work\markcite{morr}
(Morrison 1992).
 
\section{Correlation Results - Homestake Data}
 
The results of rank-order
correlations show that indeed the strongest anticorrelation
with Homestake neutrino capture rate occurs when the the
central-band surface magnetic flux is delayed.
This is shown in Figure 1 where
the Spearman correlations coefficients and significance levels
are plotted as function of time-delayed surface magnetic fields.
The strongest result is obtained when the surface field, at disk center,
is delayed by 1.4 years, i.e. a 1980 neutrino anticorrelates with
a 1981.4 surface field. Here we find a
coefficient of $-38\%$ and a significance level of $0.04\%$.
It is useful to illustrate this correlation
by plotting the Homestake data against the
delayed surface magnetic fields (Figure 2).
 
This small significance may be misleading, however,
because while the probability of an accidental $-38\%$
correlation is $0.04\%$, we correlated over many
different data sets (delays) to arrive at this optimum delay.
Statistically, if the number of data sets were large enough, we could
guarantee a small significance.
A more reliable number is perhaps the centroid of the curve of Figure 1.
>From an average of the correlation coefficients,
we find a mean delay of 0.34 years, a mean coefficient of $-23\%$, and
a significance level of $0.29\%$.
This implies a high anticorrelation probability ($> 99\%$)
between the neutrinos
and interior fields that rise to the surface in 0.34 years.
If we find the centroid of the inverse of the significance level instead
of the correlation coefficient,
we find an optimum delay of 1.0 year with
a mean correlation coefficient of $-35\%$
and a significance level of $0.46\%$.
This method places a stronger
emphasis on the very small significance levels observed around 1 year of delay.
There are, of course, many other techniques which can be
used to find the optimum delay in Figure 1, but all of them should find
strong correlations with physically reasonable delay times of 0.3-1.5 years
(as opposed to negative delays which would indicate that the surface magnetic
response precedes the interior).
 
Figure 3 shows the same correlation coefficients and delays as Figure 1
except now we include various latitude widths beyond disk center, within which
the magnetic indicators are determined.
Here we see the strongest correlations with disk center
and weaker correlations beyond (Table 1).
If we exclude the
central zone entirely (dotted line), then the probability of accidental
correlation rises dramatically (the significance level is well above 1$\%$).
This confirms the previous result (Oakley et al. 1994)
showing that the anticorrelation comes entirely
from the fields near disk center.
Numerically, the source of the
stronger anticorrelation between
neutrino captures and disk-center surface flux, as
opposed to surface flux from the whole Sun, is the
0.70-year lag between solar
maximum at 25$^{0}$ and the time when the
strongest flux arrives at the equator (not to be confused with the lag
from the solar interior).
These {\it negative} delays are clearly seen in Figure 3 and Table 1.

If neutrinos do interact with solar-magnetic fields
then perhaps information about the solar cycle itself
can also be gleaned from these latitude-dependent correlation studies.
For example, the maxima of Figure 3 for
the central and outer bands are separated by more than 0.7 years.
Might this imply that fields are held down longer near equator
due to convection zone swelling or some other unknown phenomena?
Perhaps, if these strong anti-correlations are not simply accidental but
due to a physical process.
 
\section{Correlation Results - Other Data}
 
It has been suggested that no correlation is seen between the Kamiokande
data and total sunspot number\markcite{kam1,morr}
(Hirata 1990; Morrison 1992). This can be seen
in Figure 4(a) where total sunspot number, with no delay, is plotted
against Kamiokande neutrino flux, yielding the flat line.
If we display this scatter plot
as a function of central band and time delayed
magnetic fields (Figure 4(b)), however, a
negative slope (anticorrelation) provides the best fit.
Because of the large errors and low number of data points to date, this result
may again be accidental, but anticorrelation can not be ruled out.
 
To illustrate this
for all of the available neutrino data, Figure 5 shows
neutrino data from all four experiments plotted against
the delayed magnetic data (interior fields).
The solid line shows the two year average Homestake data. Again, nothing
can be inferred from such a figure except that anticorrelation is possible.
 
It is intriguing that
anticorrelations with total solar surface indicators are weaker
than anticorrelations with proxies for interior fields along the neutrino path.
While not conclusive,
anticorrelations are present and so as more data are collected
future correlation studies should focus on these isolated field indicators,
particularly for neutrinos in the 1-7 MeV range.
 
\section{The Missing Neutrinos and Magnetic Interactions}
 
The question remains, how much of this effect is magnetic in nature?
As mentioned, it has been suggested that magnetic interactions may
account for all of the missing neutrinos, including the energy
dependence. But is this solution compatible with the observed
magnetic field variations at the surface?
 
To test the strength of the magnetic interaction,
in accounting for the missing neutrinos to first order, we can
measure the slope of the regression line in a scatter plot
of the Homestake data against the central-band flux (as shown for Kamiokande
in Figure 4).
Least-squares regression of the neutrino flux on the magnetic flux
(from $\pm$ 10$^0$ of disk center in this case)
gives at solar maximum a neutrino flux (NF) of
22\% of the expected rate (i.e. 0.3 per day, or
1.7 SNU) with a magnetic flux (MF)
of 32 units (10$^{21}$ Mx).
At solar minimum, NF = 44\% of the expected rate
and MF = 4 units. Hence it appears that
an eightfold increase in magnetic flux density
corresponds to a twofold decrease in neutrino flux.
 
If we take eightfold increase in surface field to imply an
eightfold increase in the
internal fields as well, then a simple
calculation shows, assuming both effects to be
linear, that the non-magnetic effect must remove
about half of the neutrinos from the
beam (as opposed to the 2/3 that it must
remove in the absence of any magnetic effect).
This gives an estimate for the neutrino magnetic
moment which is smaller by an order
of magnitude than that which would be
required if the magnetic field alone were removing the neutrinos.
A primarily magnetic interaction, furthermore,
seems to imply that the internal fields
near the equator swing far less dramatically
during the cycle than do the surface fields.
 
Again, large errors confuse these findings
but it does appear reasonable that if these anticorrelations are
physical in nature (and not statistical) then another process
(MSW, for example), along with direct magnetic interactions, may also
be contributing to the observed solar-neutrino deficit.

\section{Conclusions}
 
We have found that the anticorrelation between the observed Homestake
solar neutrino and Mt. Wilson magnetograph-measured
magnetic flux is greatly enhanced when we isolate the
magnetic fields these neutrinos encounter as they approach Earth.
This trend is also found in other solar neutrino data.
These magnetic fields include
not only the central latitude bands, which do strengthen
the anticorrelation, but time delayed surface fields, which proxy
the solar interior, as well.
The strongest anticorrelations are found for magnetic fields
within the central $\pm 5^{0}$ latitude band
that have been delayed by 1.4 years, but a 0.34 year delay
(taken from an average) is more likely.
This might indicate that interior fields travel to the solar surface in
this period of time.

The question of whether neutrino-magnetic interactions exist is
an interesting one and while we can not say conclusively that
we observe such an interaction, the evidence we have presented
warrants continued investigation.
As more solar-neutrino flux information is gathered,
experiments that are flavor sensitive and energy dependent would
help resolve the inconsistencies between the different experiments themselves.
The question of whether these anticorrelations are physical or statistical
in nature, however, requires data that span several solar cycles.
Future correlation studies should also focus on
the actual magnetic fields the neutrinos encounter in the solar interior.
These are most likely central-band surface fields delayed in time;
the ones that give the strongest anticorrelations here.

%%%%%%%%%%%%%%%%%%%%%%%%%%%%%% Acknowledgements
%%%%%%%%%%%%%%%%%%%%%%%%%%%%%%%%
\acknowledgments

We wish to thank R. Davis, who
supplied the Homestake data, P. McIntosh for his solar flare data, 
and J. McNeil, S. Pollock, R. J. Peterson, and the other members of
the Colorado University Nuclear Physics Laboratory
who provided many helpful conversations.
The work was supported in part by a grant from the National Science Foundation.

%%%%%%%%%%%%%%%%%%%%%%%%%%%%%%%%%% References

%%%%%%%%%%%%%%%%%%%%%%%%%%%%%%%%%%%
%%%%%%%%%%%%%%%%%%%%%%%%%%%%%%%%%%% Figures
%%%%%%%%%%%%%%%%%%%%%%%%%%%%%%%%%%%%

\clearpage

\figcaption[oak_fig1.eps]{
Results obtained
from correlations of the Homestake solar
neutrino capture rate with the time-delayed surface magnetic flux
(central-band).
The dashed line represents the Spearman rank-order
correlation coefficients while the solid line represents
the corresponding significance levels, both
plotted as a function of solar-magnetic interior-to-surface time delay.
\label{fig1}}

\figcaption[oak_fig2.eps]{
The Homestake solar
neutrino capture rate
and central-band solar-surface magnetic flux
plotted as a function of the year the data were collected.
The neutrino data are in 2-year bins and are represented
by the closed circles (and the broken line, to guide the eye).
The magnetic-flux
data are also in 2-year bins and are represented
by the crosses (and the solid line, to guide the eye).
Notice, the magnetic scale is shifted by 1-year (delayed to plot
a 1984 neutrino with a 1984 interior field, proxied by a 1985 surface field).
\label{fig2}}

\figcaption[oak_fig3.eps]{
Significance levels (rank-order) obtained
from correlations of the Homestake solar
neutrino capture rate with various latitude bands of
surface magnetic flux,
plotted as a function of magnetic interior-to-surface time delay.
Notice that the optimum delay of the outer band (straight dashed line)
precedes the central (solid line) by around 1.5 years. The horizontal line
represents the critical 1\% significance level.
\label{fig3}}

\figcaption[oak_fig4.eps]{
Scatterplots of:
(a) Kamiokande solar-neutrino flux vs. whole-disk sunspot number
(from Morrison 1992);
(b) Kamiokande vs. time-delayed (1 year), central-band surface magnetic flux.
\label{fig4}}
 
\figcaption[oak_fig3.eps]{
Solar neutrino capture rates from
Kamiokande (circles), GALLEX (diamonds), and SAGE (squares)
plotted as a function of the year the data were collected.
The neutrino data are in the time bins shown.
The 2 year averages of Homestake and
the central-band solar-surface magnetic flux
are also shown; represented
by the broken and solid lines, respectively,
as a guide to the eye.
Notice, the magnetic scale is shifted by 1-year (again delayed to plot
a 1984 neutrino with a 1984 interior field, proxied by a 1985 surface field).
\label{fig5}}
 
%%%%%%%%%%%%%%%%%%%%%%%%%%%%%%%%%%% Tables
%%%%%%%%%%%%%%%%%%%%%%%%%%%%%%%%%%%%% \newpage
\makeatletter
\def\jnl@aj{AJ}
\ifx\revtex@jnl\jnl@aj\let\tablebreak=\nl\fi
\makeatother
\begin{deluxetable}{rrrr}
\tablewidth{33pc}
\tablecaption{Correlations of the Homestake solar neutrino flux with
time-delayed surface magnetic fields from
various latitude bands. }
\tablehead{
\colhead{Band}           & \colhead{Correlation$^a$}      &
\colhead{Significance$^a$}          & \colhead{Delay$^a$}  }
\startdata
$-5^{0}<\theta <+5^{0}$ & -23\% & 0.29\%  & 0.34 yr \nl
$-10^{0}<\theta <+10^{0}$ & -23\%  &0.56\%  & 0.26 yr  \nl
$-15^{0}<\theta <+15^{0}$ & -22\%  & 1.26\%  & 0.10 yr  \nl
$-20^{0}<\theta <+20^{0}$ & -21\%  & 1.36\%  & -0.14 yr  \nl
$-25^{0}<\theta <+25^{0}$ & -17\%  & 2.88\%  & -0.28 yr  \nl

\tablenotetext{a}{From the centroid of the 
correlation coefficients (see Figure 1).}
\enddata
\end{deluxetable}

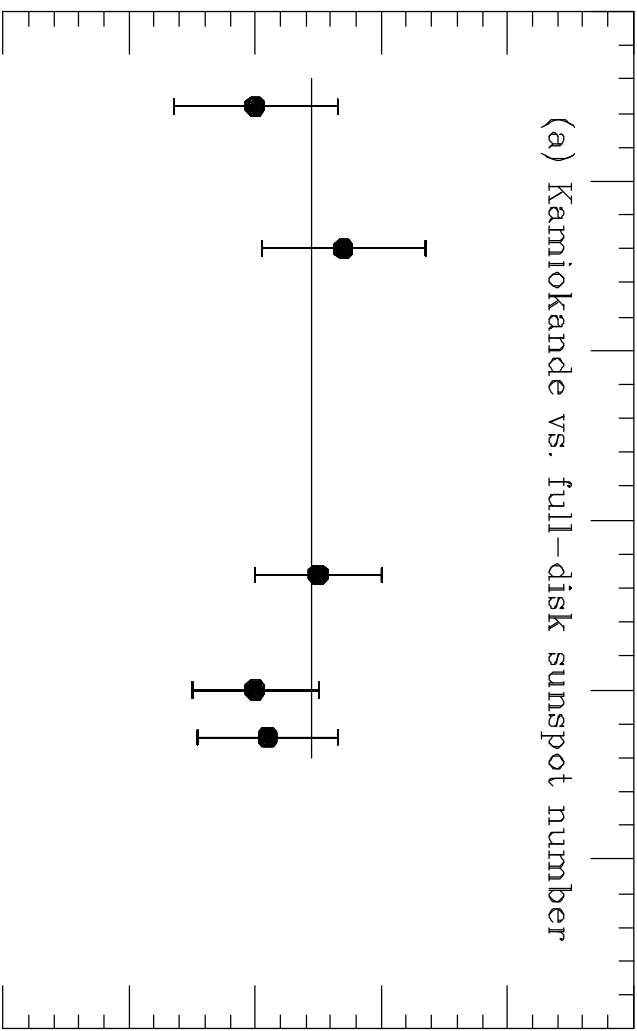
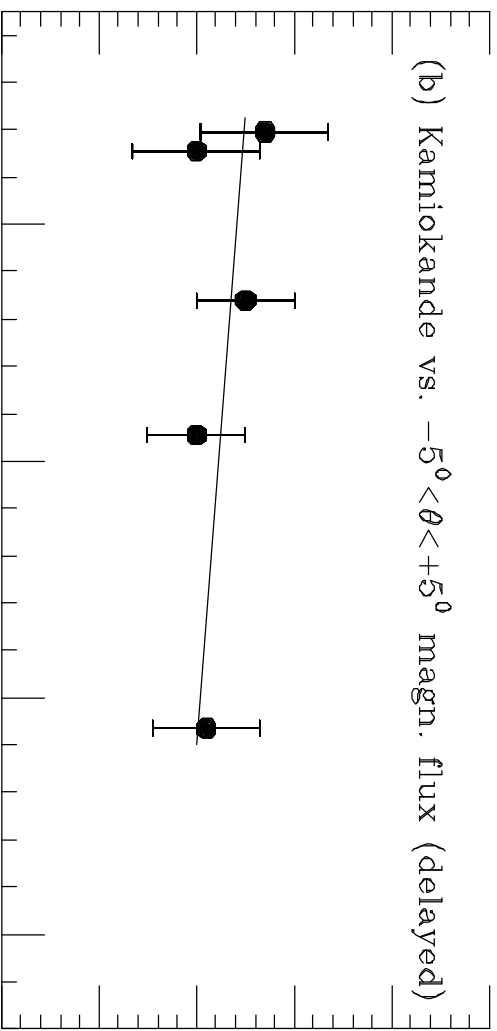
\end{document}